\documentclass[3p, preprint, times, lefttitle, twocolumn]{elsarticle}

\usepackage[english]{babel}
\usepackage[utf8]{inputenc}
\usepackage{wrapfig}
\usepackage{graphicx}
\usepackage{amsmath}
\usepackage{subcaption}
\usepackage{multirow}
\usepackage{hyperref}
\usepackage{float}
\usepackage{longtable, booktabs}
\usepackage[toc,page]{appendix}
\hypersetup{
    colorlinks=true,
    linkcolor=black,
    filecolor=magenta,      
    urlcolor=cyan,
    }

\title{Proton Beam Based Production of Positron Emitters by Exploiting the $^{27}$Al(p,x)$^{22}$Na Reaction}

\author[1]{Lisa-Marie Krug}
\ead{Lisa-Marie.Krug@frm2.tum.de}
\author[1]{Leon Cryssos}\ead{leon.cryssos@frm2.tum.de}
\author[2]{Jürgen Bundesmann}\ead{bundesmann@helmholtz-berlin.de}
\author[2]{Alina Dittwald}\ead{alina.dittwald@helmholtz-berlin.de}
\author[2]{Georgios Kourkafas}\ead{georgios.kourkafas@helmholtz-berlin.de}
\author[2]{Andrea Denker}\ead{denker@helmholtz-berlin.de}
\author[1]{Christoph Hugenschmidt\corref{cor1}}
\ead{christoph.hugenschmidt@frm2.tum.de}
\affiliation[1]{organization={Heinz Maier-Leibnitz Zentrum (MLZ), Technical University of Munich},
addressline={Lichtenbergstr. 1},
postcode={85748},
city={Garching},
country={Germany}}
\affiliation[2]{organization={Helmholtz-Zentrum Berlin für Materialien und Energie, Protonen für die Therapie},
addressline={Hahn-Meitner-Platz 1},
postcode={14109},
city={Berlin},
country={Germany}}
\cortext[cor1]{Corresponding author}

\date{March 2024}

\journal{Nuclear Instruments and Methods in Physics Research Section B}

\begin{document}

\begin{abstract}
Positron annihilation experiments on an laboratory scale depend on the supply and the availability of $\beta^+$ emitters.
Here we present the production of positron sources based on the $^{27}$Al(p,x)$^{22}$Na reaction by irradiation of Al with a 68\,MeV proton beam. 
We simulated the energy loss, range and radial scattering of the protons in Al in order to design a simple target consisting of a stack of Al discs.
Our approach allows (i) the direct use of the Al discs as positron emitters that inherently avoids wet chemical processes as usually applied in commercial production of carrier-free $^{22}$Na, (ii) the production of multiple positron sources at once, and (iii) the simple measurement of the depth and lateral distribution of $^{22}$Na.
We precisely determined the cross section of the $^{27}$Al(p,x)$^{22}$Na reaction which was found to differ from literature values particularly for proton energies between 27 and 40\,MeV. 
The activity of all nuclides produced (apart from $^{22}$Na) was shown to be negligible 15 days after irradiation.
The production of radionuclides such as $^{48}$Sc, $^{54}$Mn and $^{56}$Co can be prevented by using Al of a higher purity.
The concept presented here can easily be adapted for the production of stronger $^{22}$Na sources by increasing the proton current or/and the irradiation time. 
\end{abstract}

\bibliographystyle{elsarticle-num}

\maketitle

\section{Introduction}
\label{sec:intro}
Positron sources are essential for a wide range of applications in materials science, solid state physics and fundamental research. 
In particular, compact laboratory setups for Positron Annihilation Spectroscopy (PAS) techniques, which are non-destructive and extremely sensitive to lattice defects (see, e.g. \cite{Sch88}), cause a demand for positron sources. 
The most widely used $\beta^+$ emitter is $^{22}$Na due to its high positron yield of $90\%$, long half-life of 2.6\,years and simple handling.
In conventional PAS experiments, positrons released from $^{22}$Na with the characteristic $\beta^+$ spectrum up to the endpoint energy of 544\,keV are directly implanted into the sample material by using the compact sandwich geometry, i.e. the $^{22}$Na source is placed between two samples. 
A great advantage of $^{22}$Na is the emission of a prompt high-energy $\gamma$-quantum of 1275\,keV following the positron emission that can hence be utilized as start signal for positron lifetime measurements.

Mono-energetic positron beams of \textit{high} intensity, i.e. facilities providing more than $10^7$ moderated positrons per second \cite{Hug10a}, are generated by pair production at large scale facilities such as electron linacs \cite{Suz97a, Wad13, Kra08}  or research reactors \cite{Vee01, Hug08b}. 
For the operation of laboratory positron beams, however, $^{22}$Na is indispensable. 
Here the positron source is usually combined with solid neon \cite{Mil86} or tungsten \cite{Str01} moderators to generate mono-energetic positrons. 
Such setups, which typically provide intensities of $10^4-10^6$ moderated positrons per second, are used for numerous experiments such as depth dependent PAS \cite{Coleman2000, Gid06}, surface studies \cite{Kaw00, Str01, Muk10}, positronium experiments \cite{Mar05, Cas06}, or trap-based experiments \cite{Mur92, Gre02b}.

In principle, $^{22}$Na can be produced by irradiation of  an Al or Mg target with protons, deuterons or $\alpha$ particles of appropriate energy provided by the particle accelerators.
The maximum  source activity is limited by the irradiation time, target cooling and the available ion current.
After chemical separation, small positron sources are prepared by drying $^{22}$NaCl or $^{22}$Na$_2$CO$_3$ from its solution onto thin foils or in small capsules.

The main drawback of all these positron experiments, however, is the very limited availability of $^{22}$Na.
At present, there is only one supplier worldwide, iThemba LABS \cite{vdW04}, that produces and distributes $^{22}$Na sources. 
In this research institution $^{22}$Na is chemically separated from a magnesium target, which has been irradiated with protons, and then filled into a metallic capsule.
For this reason, alternative routes for the generation and supply of positron emitters -- especially $^{22}$Na sources -- are urgently needed.

In this paper, we present the  basic concept for producing $^{22}$Na sources by irradiating an optimized Al target with 68\,MeV protons provided by the  cyclotron at the Helmholtz-Zentrum in Berlin \cite{Den23}.
For target design, simulations are conducted to investigate the range of protons and the radial scattering of the beam as well as the energy loss in Al.
The target consisting of a stack of Al discs enables the production of multiple positron sources at once and allows us to easily measure the depth distribution of $^{22}$Na.
The approach of the $^{22}$Na production in thin Al discs, which can be directly used as positron emitters, is expected to be simple and efficient especially by eliminating wet-chemical processes. 
The depth-dependent activity of $^{22}$Na determined by measuring the count rate of each Al disc will be compared to the depth distribution calculated from the simulated local proton energy and literature values for the $^{27}$Al(p,x)$^{22}$Na reaction cross section. 
Finally, $\gamma$-spectra of the irradiated target are recorded in order to identify the nuclides produced and to determine their activity.

\section{Simulations and Target Design}
\label{chapter:sampledesign}
To determine the dimensions of the target, a simulation was conducted using the program \textit{The Stopping and Range of Ions in Matter (SRIM)} \cite{srim}. We simulated the irradiation of a block of pure aluminum by a 68\,MeV proton beam by using 100,000 particles. 
The simulation data is used to estimate the penetration depth and radial distribution of the protons as well as their energy loss in the target.

\begin{figure}[t!]
	\centering
	\includegraphics[width=0.5\textwidth]{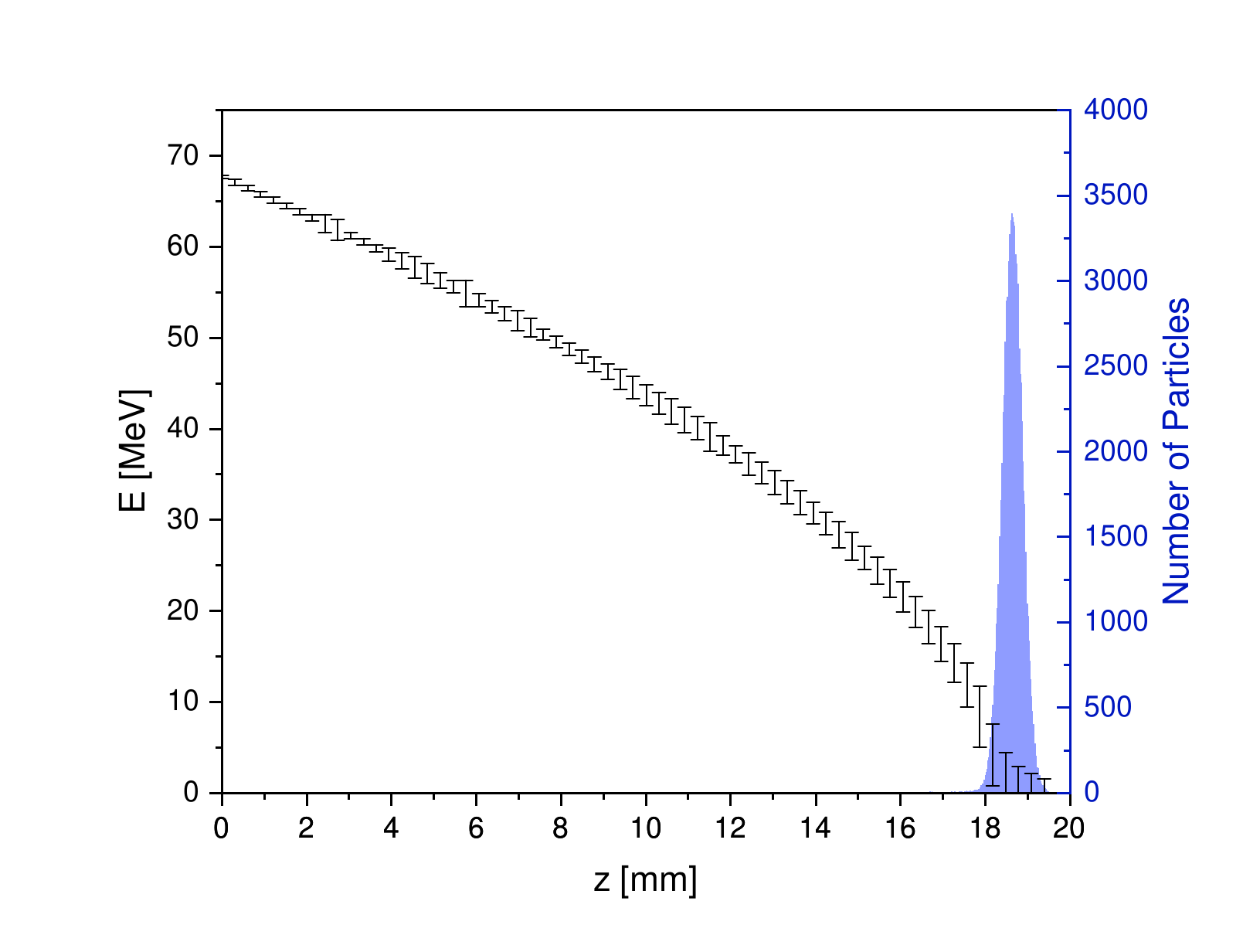}
	\caption{Implantation of 68\,MeV protons in Al: Distribution of the protons stopped (blue) and mean proton energy $E$ as a function of depth $z$ in aluminum (black); the errorbars correspond to the standard deviation of the 100,000 simulated protons. 
 }
	\label{fig:energyerror}
\end{figure}

\begin{figure}[t!]
	\centering
	\includegraphics[width=0.5\textwidth]{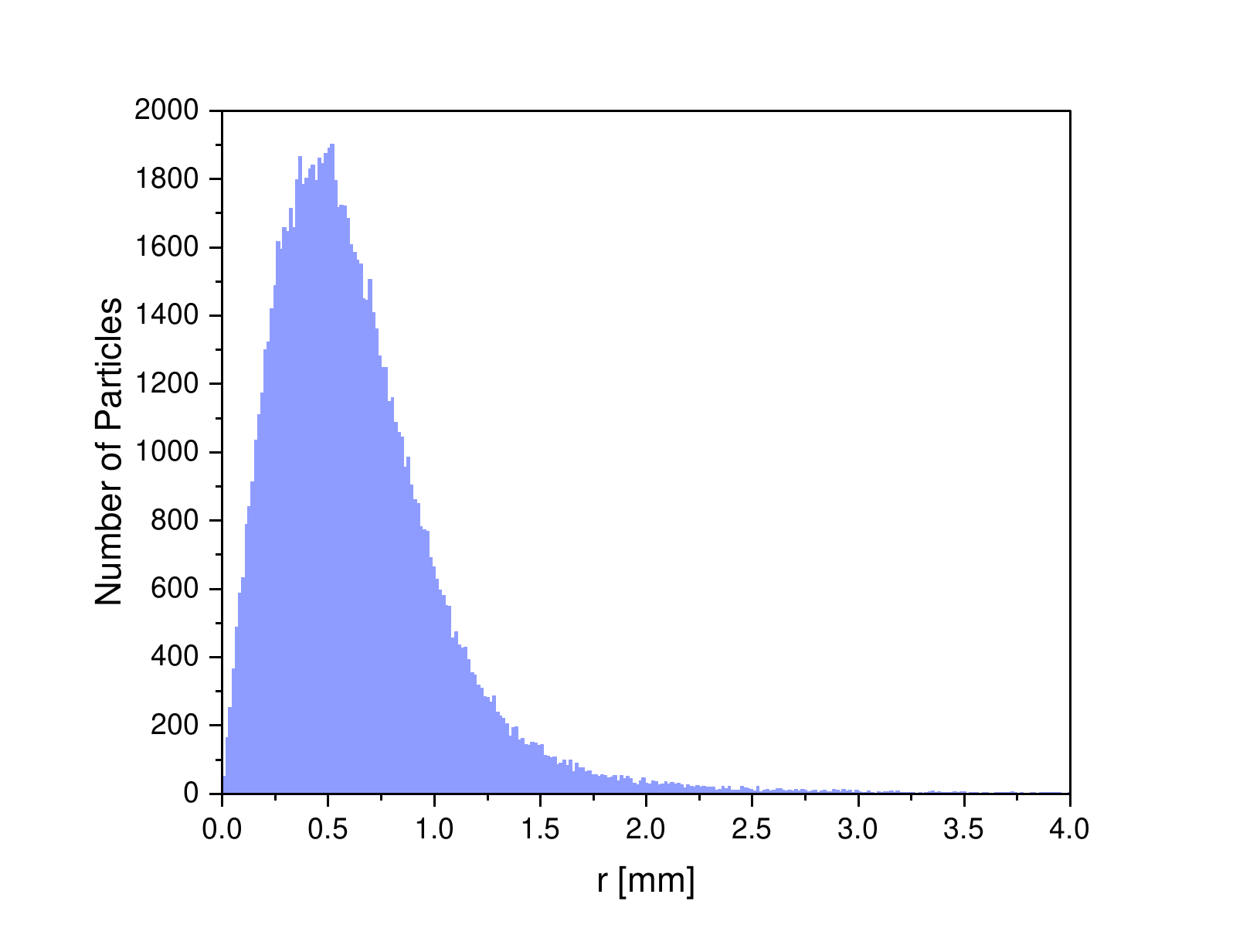}
	\caption{Distribution of radial scattering using SRIM data with 100,000 protons. Their mean radial scattering is  about 0.65\,mm.}
	\label{fig:dataradius}
\end{figure}

The average penetration depth at which the protons are stopped is about 18.6\,mm as shown in Figure\,\ref{fig:energyerror}.
Since the depth distribution of the produced $^{22}$Na depends on both the energy dependent cross section for the $^{27}$Al(p,x)$^{22}$Na reaction \cite{crosssection} and the local proton energy we simulated the energy loss in aluminum as well. 
In order to calculate the depth-dependent average kinetic energy of the protons the depth was divided into 66 bins (one bin for each aluminum disc, see below).
The resulting mean proton energy $E$ as function of depth in the target is also plotted in Figure\,\ref{fig:energyerror}.
The radial distribution of protons at the end of their path is shown as histogram in Figure \ref{fig:dataradius}. 
It exhibits a mean radial scatter of 0.65\,mm and only very few protons scatter more than 2\,mm. 

\begin{figure}[t!]
	\centering
	\includegraphics[width=0.5\textwidth]{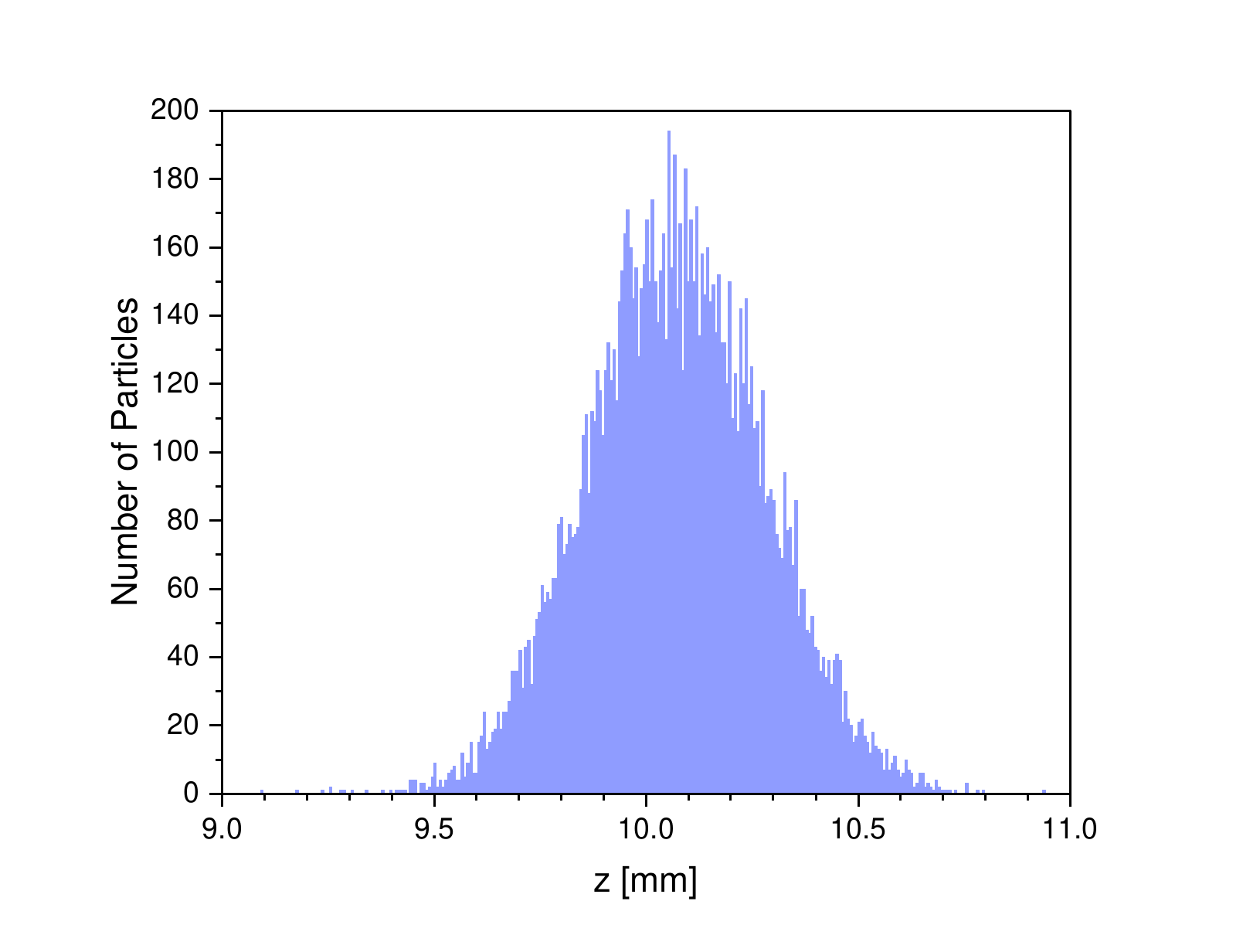}
	\caption{Histogram showing the distribution around the mean depth of 10.07(01)\,mm at which the protons reach an energy of 43.9-44.1\,MeV. }
	\label{fig:datamax}
\end{figure}

For estimating the depth distribution and the maximum of $^{22}$Na production we combine the depth dependent proton energy (see Figure\,\ref{fig:energyerror}) and the energy dependent cross section by using tabulated values \cite{tabCS} (see also black line plotted in Figure\,\ref{fig:correctedcs}).
Since  the maximum of the cross section is found to be at 44\,MeV we filtered proton energies in the range of 43.9 to 44.1\,MeV from the rest to determine their depth distribution and the exact position of the maximum.
The corresponding depths of these protons with 44\,$\pm$0.1\,MeV energy are displayed as a histogram in Figure \ref{fig:datamax}. Accordingly, the maximum of the $^{22}$Na activity produced will be expected at a depth of 10.07(1)\,mm.\par

The shape of the proton beam at the target station is estimated to be elliptical with major axes of 3 and 6\,mm. 
Therefore, the final radial activity distribution is governed by the spot size of the proton beam and not by the lateral scattering of the protons (see Figure\,\ref{fig:dataradius}).
By taking into account potential beam displacement and some space for the target holder, the geometry of the Al target was designed with a 20\,mm diameter and a total thickness of 20\,mm according to the simulated maximum proton penetration depth of about 18.6\,mm.
We decided to use a stack of thin Al discs instead of an Al block in order to enable a simple experimental determination of the achieved depth distribution of $^{22}$Na by simply measuring the activity of the individual Al discs.
By considering the simulation result for the depth dependent proton energy (Figure\,\ref{fig:energyerror}) and in order to minimize the number of measurements we chose a reasonable disc thickness of 0.3\,mm.
(Note that for the application of $^{22}$Na as positron emitters in PAS a much thinner disc of about 0.1\,mm or below is recommended to minimize positron loss due to self absorption in the Al disc itself.)

\begin{figure}
	\centering
        \subfloat[\label{fig:samplesheets}]
	{\includegraphics[width=0.23\textwidth, height=4cm]{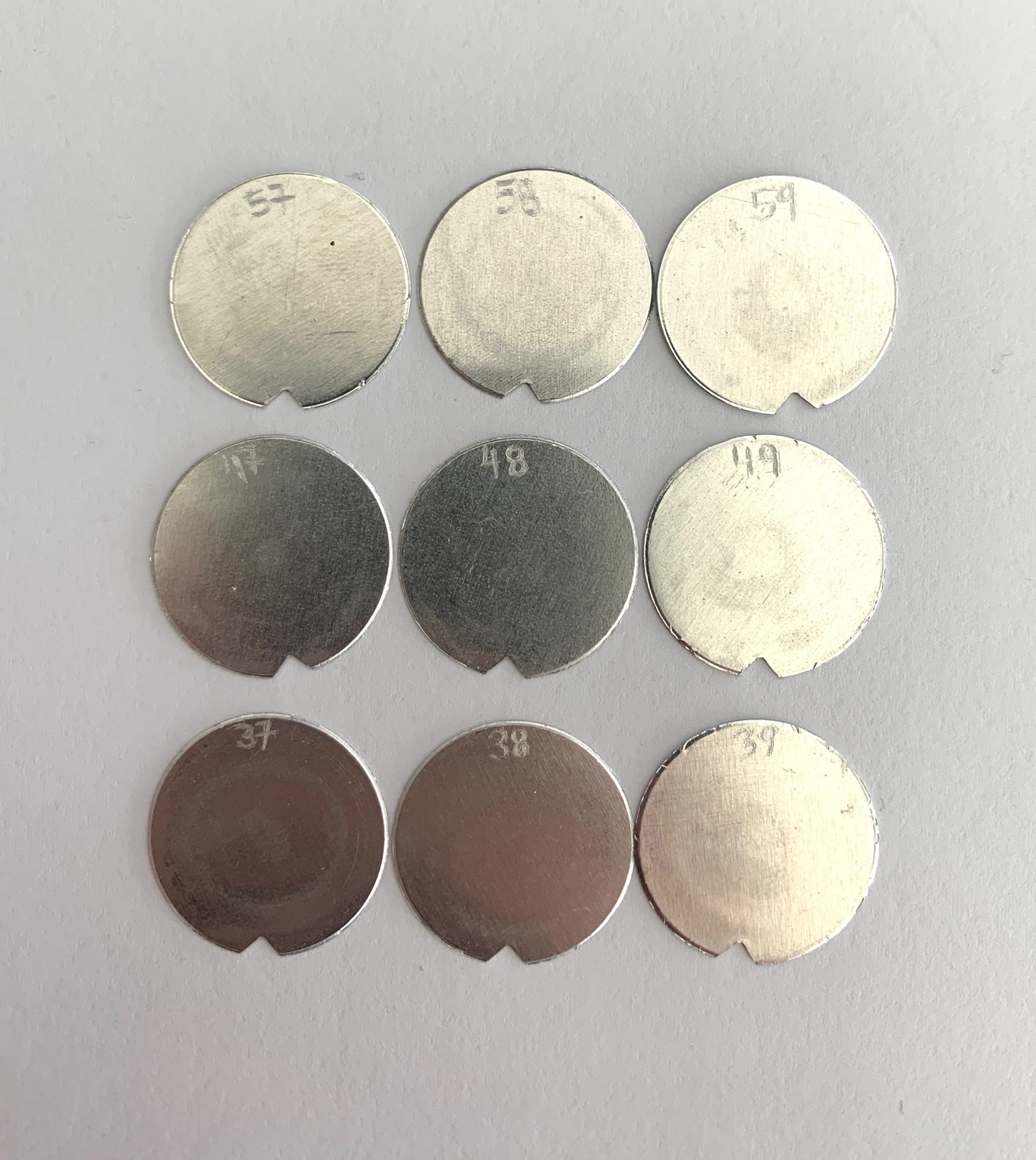}}\hfill
	\subfloat[\label{fig:sampletopf}] {\includegraphics[width=0.23\textwidth,height=4cm]{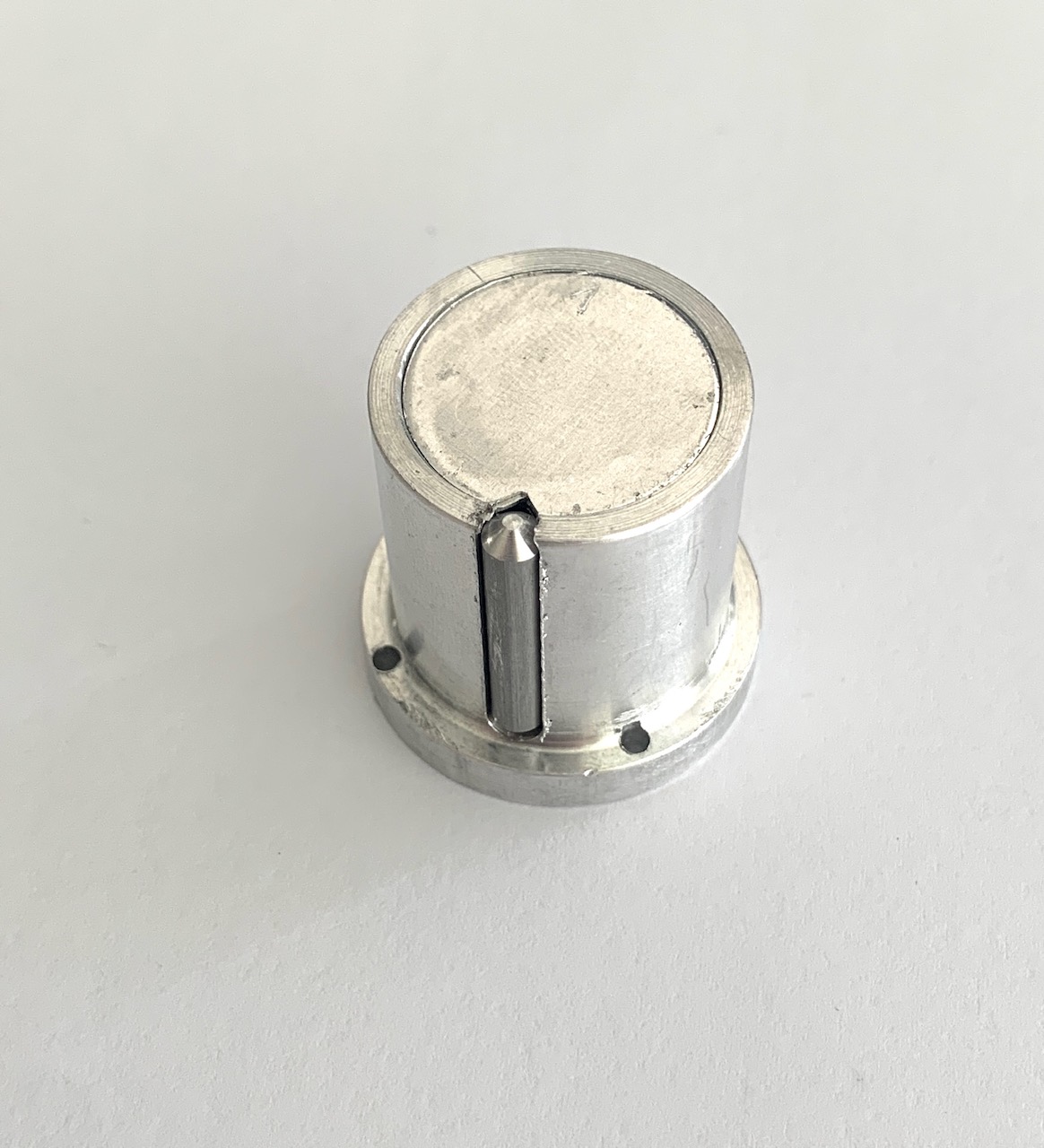}}\hfill
	\subfloat[\label{fig:sampledeckel}]
	{\includegraphics[width=0.23\textwidth,height=4cm]{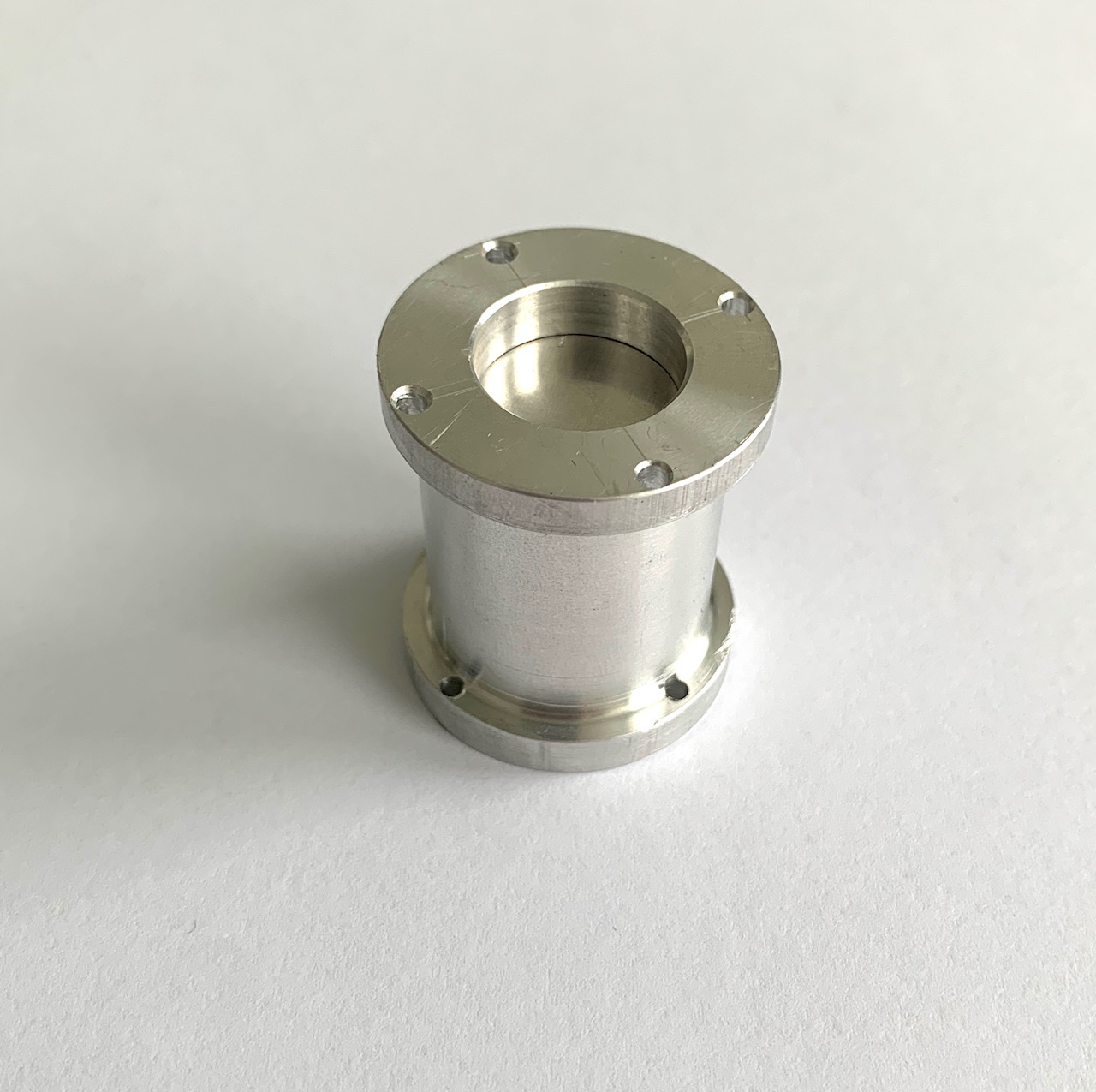}}\hfill
	\subfloat[\label{fig:sampleholder}]
	{\includegraphics[width=0.23\textwidth]{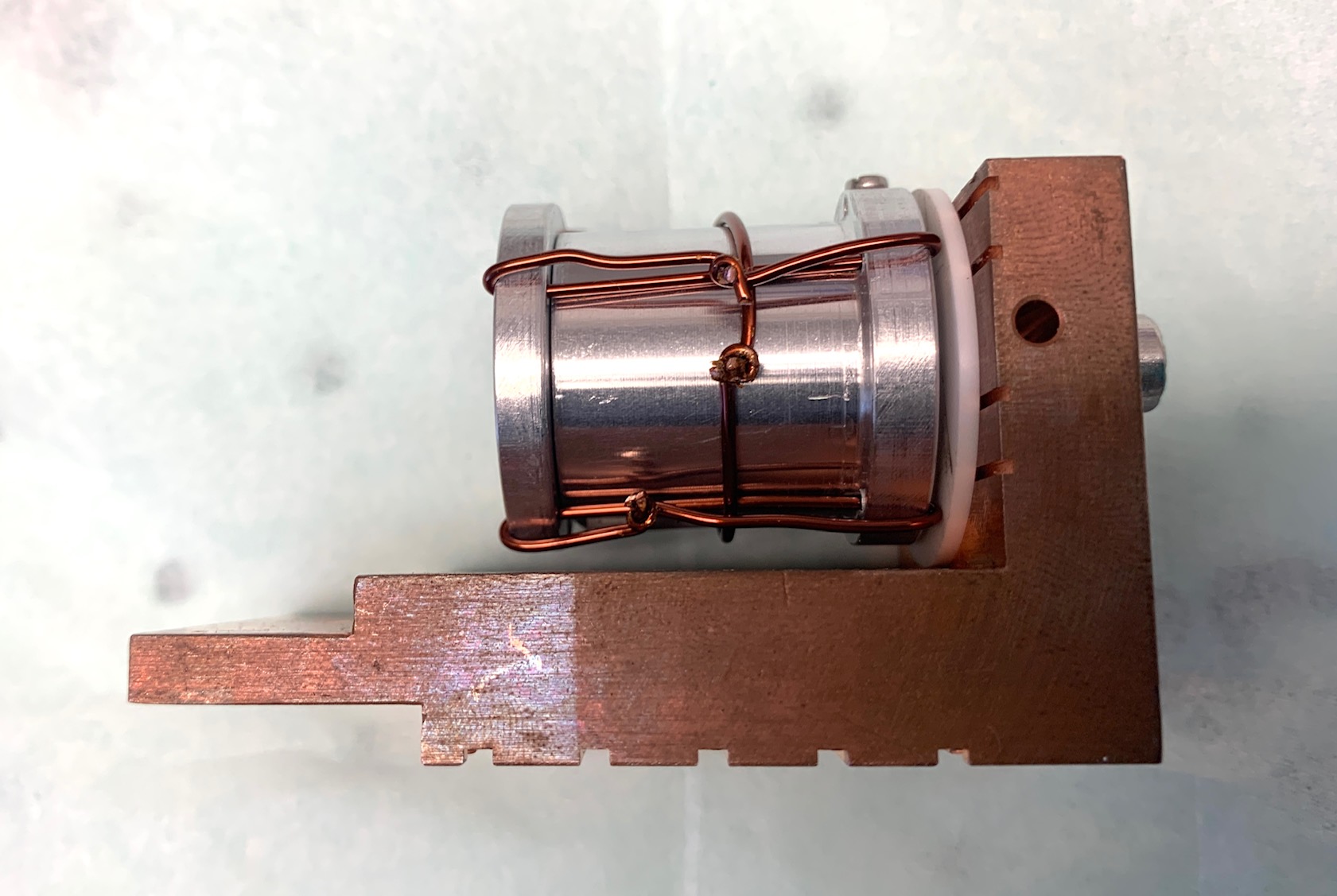}}
	\caption{Design of the Al target for $^{22}$Na production: a) Al discs 0.3\,mm thick with a diameter of 20\,mm and an indentation for rod placement and numbering to keep track of their sequence. b) Al discs stacked (19.8\,mm high) into a cylindrical case with disc number one at the top. A rod prevents the movement of the discs. c) Case with discs inside and lid on top. The holes are used to press the lid onto the cylinder. d) Cylindrical case and lid held together using copper wire and mounted onto the sample holder. The Al target is  isolated to allow the measurement of the proton current.} \label{fig:sample}
\end{figure}

As shown in the photos in Figure\,\ref{fig:sample} the final sample setup consists of a stack of 66 circular, 0.3\,mm thick aluminum plates (discs) with a diameter of 20\,mm that are placed in a cylindrical tube made of aluminum. The 20\,mm high stack is held together by a lid, which is connected to the bottom of the cylinder by copper wire. A number is carved into each disc to document the order of the discs inside the cylinder and a small Al rod is used at the bottom to prevent the discs from twisting. The full, closed cylinder weighs 48\,g and is placed onto the copper holder of a Faraday cup provided by the HZB. Active target cooling is not required as the heat load deposited by the proton beam amounts only to 6.8\,W. The number of protons hitting the target can be determined by measuring the current since the cylinder is isolated from the holder. 

\section{Results and Discussion}
\label{chapter:results}
\subsection{Proton irradiation, $^{22}$Na production and activities}
Before the final irradiation a test run with a proton beam of 1\,nA for 60\,s was conducted in order to detect any further produced nuclides and to check the dose rate at the target after the experiment. 
Immediately after irradiation a dose rate of 374\,$\mu$Sv/h was measured at a distance of 6\,cm while the target was still installed at its position in the beam line. The measurement was repeated in 30 minute intervals to estimate the lifetimes of the short-lived nuclides. The fit to the data results in a mean lifetime of about 0.5\,h summarizing all short-lived nuclides. Considering the other (p,x) reactions in aluminum \cite{crosssection} these nuclides are most likely $^{27}$Si, $^{23}$Mg and $^{18}$F with half lives of $4.15(4)$\,s, $11.305(5)$\,s and $109.77(5)$\,min respectively \cite{chart}.

\begin{figure}[h!]
	\centering
	\includegraphics[width=0.5\textwidth]{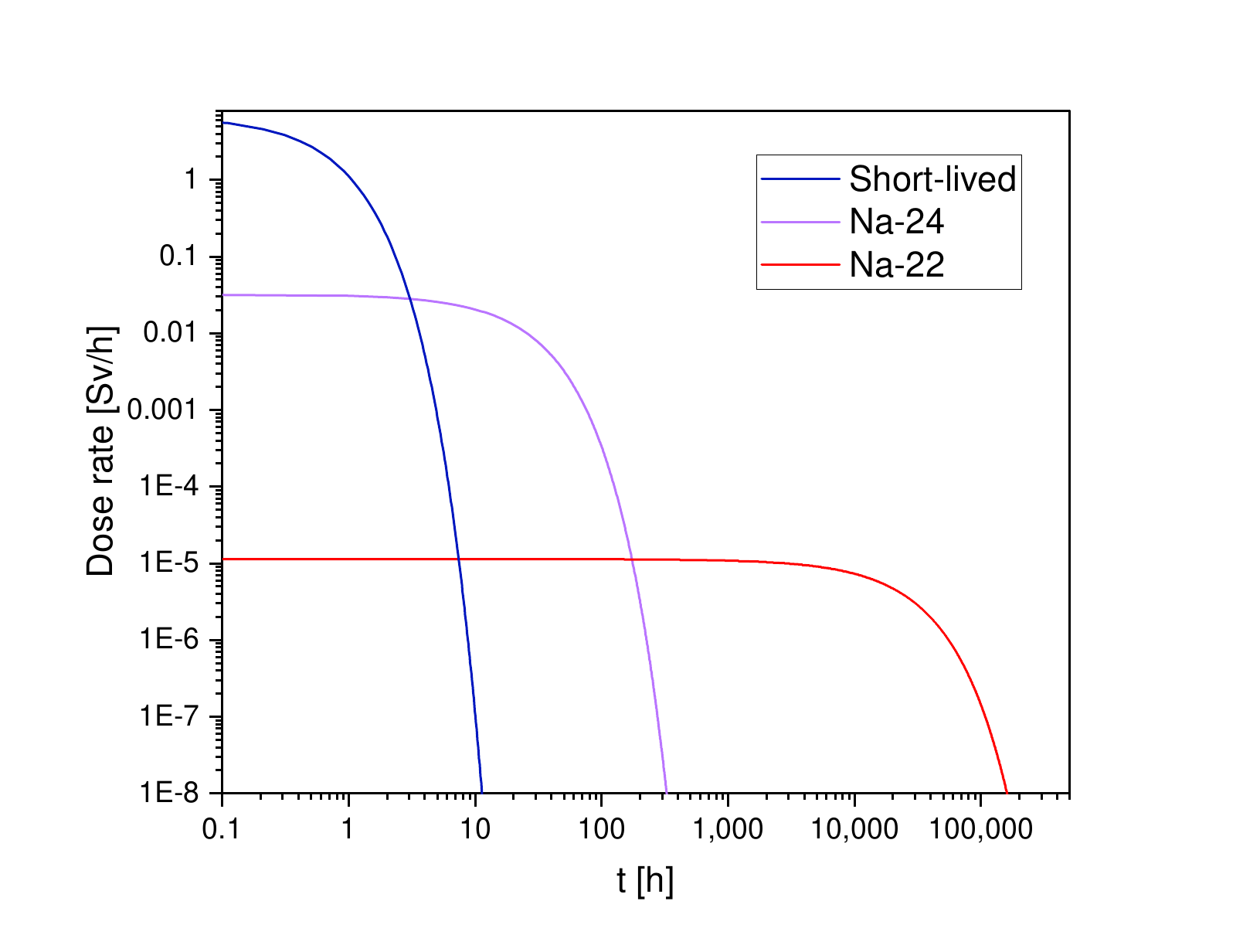}
	\caption{Dose rate after final irradiation of short-lived nuclides, $^{24}$Na and $^{22}$Na as function of time $t$. It takes about 10 hours for the short-lived nuclides and \mbox{15 days} for $^{24}$Na to decay to negligible amounts.}
	\label{fig:dosisleistung}
\end{figure}

The dose rate and activity of $^{22}$Na and $^{24}$Na after the final irradiation (proton current of 100\,nA for 3 hours) can be estimated using the data from the test run. 
For the unknown nuclides with a short lifetime the dose rate is scaled analogously. For comparability to the short-lived nuclides, the activities of $^{24}$Na and $^{22}$Na were converted to dose rates. The dose rate over time of all nuclides produced is shown in Figure \ref{fig:dosisleistung}. Note that the short-lived nuclides decay after 10 hours, already reducing the total dose rate significantly. After 350 hours, i.e. about 15 days, $^{24}$Na will have decayed too and only $^{22}$Na remains. Therefore, to minimize radiation exposure the sample is retrieved at least ten days after the irradiation is concluded.

During the final irradiation further long-lived nuclides aside from $^{22}$Na were produced mainly from impurities in the aluminum target material. All nuclides produced and their properties are summarized in Table\,\ref{tab:finalnuclides}.
$^7$Be is produced in a proton-induced spallation reaction. The heavier radionuclides are produced from heavy element impurities during reactions initiated by either the incoming proton beam directly or secondary particles emitted in other reactions.

\begin{table}[t!]
	\centering
        \scalebox{0.65}{
	\begin{tabular}{l|l|l|l|l|l}
		\textbf{Nuklide}  & \textbf{Activity [Bq]} & \textbf{Proportion} & \textbf{Half Life} & \textbf{Decay} & \textbf{$\gamma$-Energy} \\
		 & & & \textbf{[Days] \cite{chart}} & \textbf{Mode} \cite{chart}& \textbf{[keV] \cite{chart}} \\
		\hline
		$^7$Be & $2.93(21) \cdot 10^4$ & 14.24\% & 53.22 & electron & 477 \\
			 &  &  &  & capture & \\ \hline
		$^{22}$Na & $1.40(06) \cdot 10^5$ & 68.04\% & 912.5 & $\beta^+$ & 511 \\
		 &  &  &  &  & 1274.5 \\ \hline
		$^{24}$Na & $1.30(33) \cdot 10^2$ & 0.06\% & 0.625 & $\beta^-$ & 1369 \\
		 &  &  &  &  & 2754 \\ \hline
		$^{46}$Sc & $1.95(95) \cdot 10^2$ & 0.09\% & 83.79 & $\beta^-$ & 889 \\
		 &  &  &  &  & 1120 \\ \hline
		$^{48}$Sc & $3.30(13) \cdot 10^3$ & 1.60\% & 1.819 & $\beta^-$  & 983.5 \\ 
		 &  &  &  &  & 1312 \\ \hline
		$^{51}$Cr & $(8.88\pm1.38) \cdot 10^3$ & 4.31\% & 27.5 & electron & 320 \\ 
		  &  &  &  & capture & \\ \hline
		$^{52}$Mn & $5.18(16) \cdot 10^3$ & 2.52\% & 5.591 & $\beta^+$ & 511 \\ 
		 ~ & ~ & ~ & ~ & ~ & 935.5 \\
		 ~ & ~ & ~ & ~ & ~ & 1434 \\ \hline
		$^{54}$Mn & $1.78(14) \cdot 10^3$ & 0.86\% & 312.2 & $\beta^+$ & 511 \\ 
		 ~ & ~ & ~ & ~ & ~ & 835 \\ \hline
		$^{56}$Co & $2.46(09) \cdot 10^3$ & 1.20\% & 77.236 & $\beta^+$ & 511 \\
		~ & ~ & ~ & ~ & ~ & 847.7 \\
		~ & ~ & ~ & ~ & ~ & 1238 \\ \hline
		$^{58}$Co & $4.53(72) \cdot 10^2$ & 0.22\% & 70.86 & $\beta^+$ & 511 \\ 
		 ~ & ~ & ~ & ~ & ~ & 810 \\ \hline
		$^{67}$Ga & $1.31(27) \cdot 10^3$ & 0.63\% & 3.2617 & electron & 93.3\\
		~ & ~ & ~ & ~ & capture & 184.5 \\
		~ & ~ & ~ & ~ &  & 393.5 \\	
	\end{tabular}}
	\caption{Radionuclides found in the target 12.5 days after final irradiation}
	\label{tab:finalnuclides}
\end{table}

\subsection{Depth distribution of $^{22}$Na}
In order to determine the activity of $^{22}$Na produced we measured the count rate of the 1275\,keV $\gamma$-quanta emitted by the de-excitation  of the excited $^{22}$Ne state following the $\beta^{+}$ decay of $^{22}$Na (probability 90\,\%).
The count rate of the 511\,keV annihilation peak could not be used because other nuclides also produce positrons, which adds to the signal inconsistently.

For each Al disc, a $\gamma$-spectrum was recorded using a High Purity Germanium (HPGe) detector with an efficiency of 30\% and an energy resolution of 0.22\% at 1275\,keV. The distance between sample and the detector was approximately 10\,cm and each disc was measured for five minutes resulting in net counts up to 26,000 in the 511\,keV photo peak and 5,300 in the 1275\,keV peak. The measurement duration was increased for the last 15 discs due to the low amount of counts. To determine the count rate of each disc, the net counts found in the 1275\,keV peak were divided by the measurement time. 

\begin{figure}[h!]
	\centering
	\includegraphics[width=0.5\textwidth]{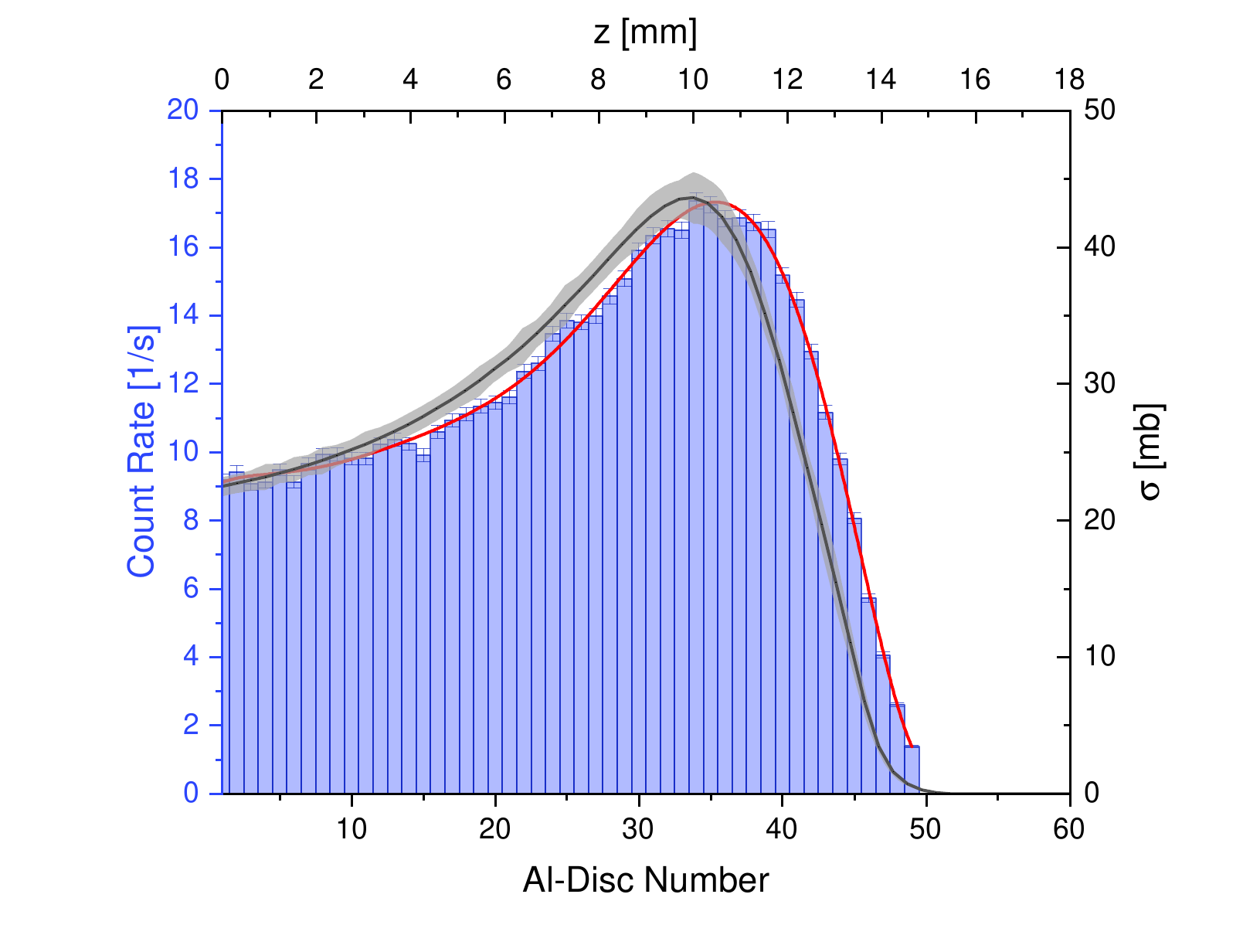}
	\caption{Count rate of the 1275\,keV photo peak of $^{22}$Na as a function of the disc number, i.e.\,target depth (blue); the depth distribution of $^{22}$Na is proportional to the measured count rate. 
	The predicted $^{22}$Na distribution is expected to follow the depth dependent cross section calculated from its energy-dependent values taken from literature \cite{crosssection} and SRIM simulation (black line; error margin as grey shading). The maximum of simulation and measurement match but a significant offset of about 1.5\,mm is observed.
	The fit to the experimental data (red line) is obtained by the model explained in Section\,\ref{Sec:New_Sigma}.}
	\label{fig:tiefenprofil}
\end{figure}

Figure\,\ref{fig:tiefenprofil} shows the depth profile of the intensity of the 1275\,keV photo peak obtained from measuring the discs, which is hence equivalent to averaging the count rate over a depth of 0.3\,mm. 
Since the measured count rate is proportional to the $^{22}$Na activity the curve calculated shows the depth distribution of $^{22}$Na produced.
Thus the integral over this curve corresponds to the total activity $^{22}$Na measured for the whole target that amounts to 140\,kBq (see Table\,\ref{tab:finalnuclides}).
It can be seen that disc number 34, which corresponds to a depth of 9.9 to 10.2\,mm, exhibits the highest activity. 
Its activity can be easily determined by comparing the count rates of the most active disc and the whole target as the measurement time of the spectra was 5 minutes and the orientation of the objects were similar. 
This yields a $^{22}$Na activity of 4.62(23)\,kBq in disc no.\,34.

For comparison, in Figure\,\ref{fig:tiefenprofil} we also plot the depth dependent cross section calculated from its energy dependent values taken from literature (see black line in Figure\,\ref{fig:correctedcs} and Table \ref{tab:NEWcrosssection}) and the simulated proton energies. The error is dominated by the accuracy of the tabulated literature values for the cross section and is shown as grey shaded area.
The measurements are in qualitative good agreement with the shape of the cross section curve from the calculated values. 
The simulation predicted the maximum to be at 10.07\,mm, which lies within disc no.\,34 as measured. 
Note that a general slight shift of the experimental data points of about 1.5\,mm towards greater depth is observed.
This deviation, however, cannot be explained by channeling, i.e. preferential proton implantation along high-symmetry crystal directions, since this effect is expected to play a negligible role in the poly-crystalline target material.
Although the exact kinetic energy of the proton beam was not specifically determined for the present irradiation experiment, uncertainties of the beam energy are estimated to be small in order to explain the observed effect.
(At a previous experiment, the maximum beam energy was determined to be less than 0.4\% higher compared to the nominal proton energy of 68\,MeV delivered by the cyclotron; this would correspond to an increase of the proton penetration depth in Al by only about 0.1\,mm.)
For this reason, the main cause of the observed shift is rather attributed to the uncertainty of the cross section values taken for the simulation.

\subsection{Lateral distribution of $^{22}$Na}
\begin{figure}[h!]
	\centering
	\includegraphics[width=0.5\textwidth]{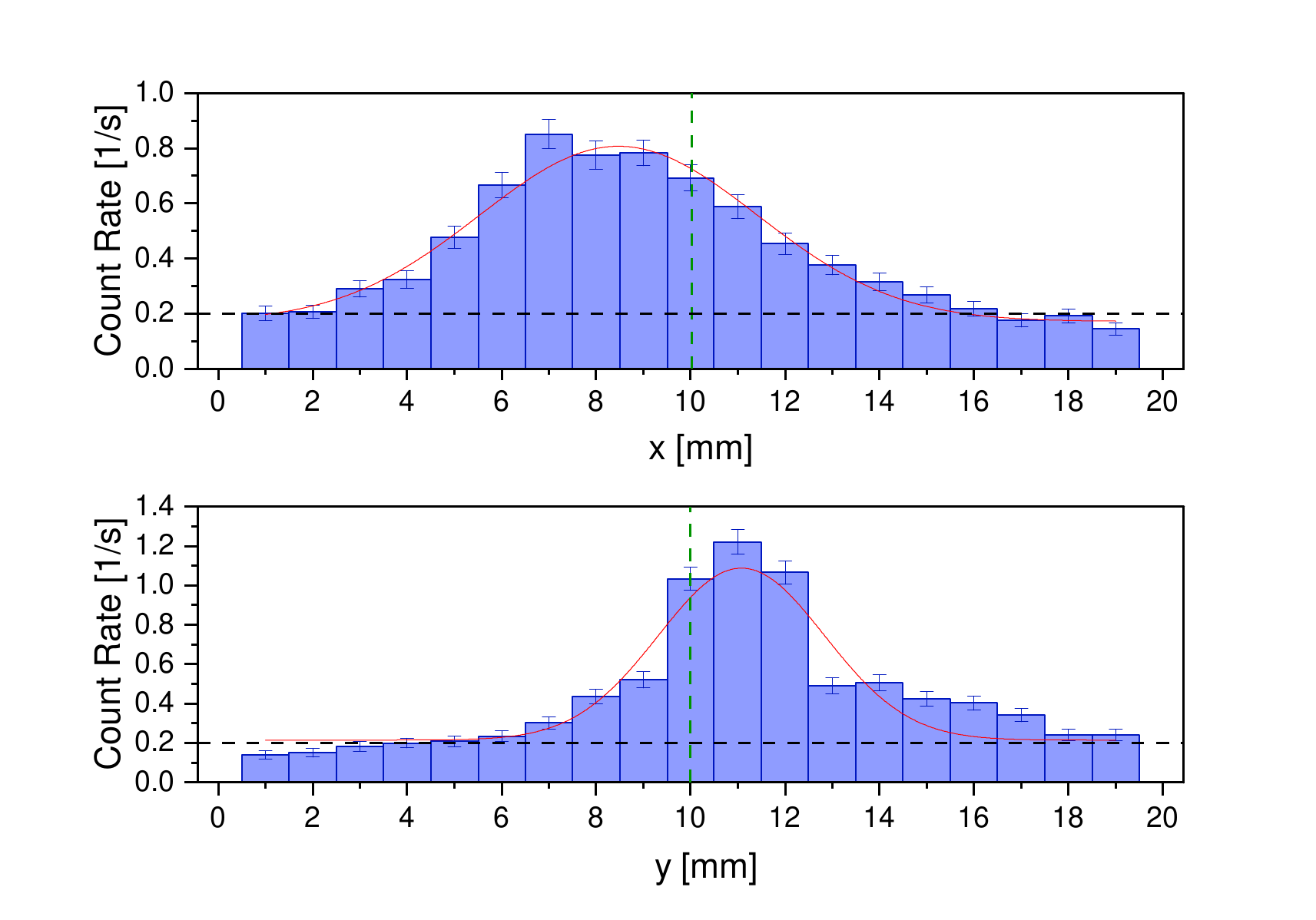}
	\caption{Horizontal (top) and vertical (bottom) measurement of the count rate of the most active disc no.\,34.
 The background count rate was 0.2\,s$^{-1}$ (black dashed lines) and the center of the target was at (x,y)=(10\,mm, 10\,mm) (green dashed lines).
  The Gaussian fits result in maxima at 8.39(13) and 11.07(16)\,mm with widths of 7.50(09) and 4.13(22)\,mm (FWHM) in x and y direction respectively. 
  }
	\label{fig:radial}
\end{figure}

The radial distribution of the count rate in the most active disc no.\,34 was measured by using a tungsten collimator with a 1\,mm broad slit and moving the disc over the slit in 1\,mm steps. At each position a spectrum was taken for five minutes and the counts in the 1275\,keV peak were used to calculate the count rate. 
The projection of the radial profile was measured in both the x and y direction. Figure \ref{fig:radial} shows the measured count rate  in horizontal (x) and vertical (y) direction relative to the direction of the proton beam.  A Gaussian profile was fitted to the data to determine the full width half maximum (FWHM) and the position of the maximum. 
As result we find the centers of the profiles at 8.39(13) and 11.07(16)\,mm with widths of 7.50(09) and 4.13(22)\,mm (FWHM) in x and y direction respectively. 

As suggested by both lateral measurements, the target was not hit in the center but at a position slightly shifted to the top left. 
The deviation of the maximum from the center is -1.61(13) and 1.07(16)\,mm in the x and y direction respectively. This deviation can be explained by the symmetry axis of the target, which may not have been fully aligned with the direction of the proton beam during irradiation. 
A possible tilting of small angles between the target and the proton beam can also cause a shift of the beam spot especially in the rear discs.

The cross section of the proton beam is assumed to be elliptical with half axes of 6 and 3\,mm in x- and y-direction respectively.
When considering the beam spot broadening due to scattering (see simulation result in Figure\,\ref{fig:dataradius}) the measured full width half maxima are in agreement with the dimensions of the elliptical beam.
However, a count rate significantly above the background level (above 13 and below 9\,mm in y-direction) might indicate the presence of a beam halo. 

\subsection{Revised cross-section values for $^{22}$Na production}
\label{Sec:New_Sigma}
In the following we want to derive a reliable model to explain the deviation observed for the measured and the calculated depth dependent distribution of $^{22}$Na in the aluminum target as shown in Figure\,\ref{fig:tiefenprofil}.
For this, we will have a closer look on the cross-section for the $^{27}$Al(p,x)$^{22}$Na reaction as function of proton energy.

\begin{figure}[h]
	\centering
	\includegraphics[width=0.5\textwidth]{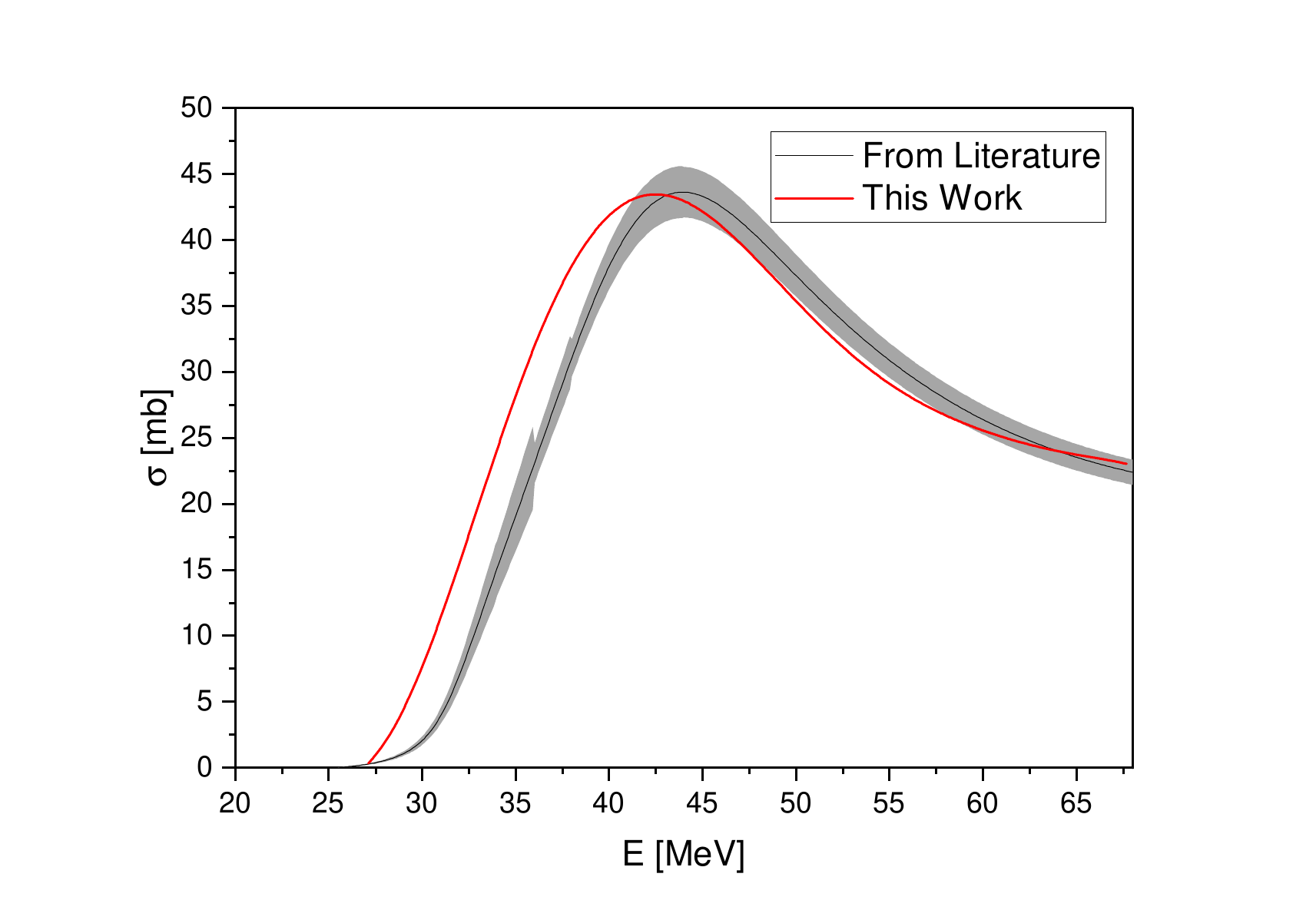}
	\caption{Cross section for the $^{27}$Al(p,x)$^{22}$Na reaction: values taken from literature with error margin (black line with grey shading) \cite{crosssection} and new values obtained by fitting the experimental data as shown in Figure\,\ref{fig:tiefenprofil} (red line).}
	\label{fig:correctedcs}
\end{figure}

Apparently, the calculated cross-section based on values taken from literature \cite{crosssection} overestimates the $^{22}$Na activity at depth $<$10\,mm, i.e.\,at higher proton energies and underestimates it at larger depths, i.e.\,at lower proton energies.
In order to achieve a better agreement, we vary the  \textit{energy-dependent} cross section --starting from the literature values-- and re-calculate the  \textit{depth-dependent} reaction cross section. 
In such a way we run the optimization until the best fit to the experimental data is obtained as shown in Figure\,\ref{fig:tiefenprofil}.
The input data for this fit (see also Table \ref{tab:NEWcrosssection} in Appendix \ref{appendix:crosssection}), i.e. the cross section as function of energy is plotted in Figure\,\ref{fig:correctedcs} alongside with the values taken from literature.
In particular a significant difference  is seen in the energy range of 27-40\,MeV whereas at larger proton energy the deviation of the new values is still compatible with the error margin of the literature values.

\section{New Route for Production of Positron Emitters?}
For the application of the activated Al discs as positron emitters their thickness should be low in order to minimize positron loss due to positron self absorption.
Taking into account the stopping power for positrons with an energy of 215\,keV (mean energy of the $\beta^+$ spectrum of $^{22}$Na \cite{energybeta}) of 2.112\,MeV cm$^2$/g taken from the standard reference database from NIST \cite{stoppingpower} and the mass density of Al (2.7\,g/cm$^3$) a thickness of 50\,$\mu$m seems to be reasonable.

The irradiation scheme presented here has the charm of simultaneous production of numerous positron sources with a single target setup.
In addition, our approach avoids wet chemical processes, which are used in commercial production of carrier-free $^{22}$Na sources, by using the activated aluminum discs as positron emitters. 
If required, a protective layer can be prepared by targeted oxidation of the aluminum discs for improving the robustness of the surface.
Furthermore, the production of other radionuclides among $^{48}$Sc, $^{54}$Mn and $^{56}$Co, aside from $^7$Be, can easily be prevented by using aluminum of a higher purity.

For positron annihilation techniques it is desirable to provide positron sources with activities in the order of 100\,kBq. 
Stronger sources than generated in our experiment presented here can easily be produced by increasing the proton flux or/and the irradiation time.
For instance, irradiating the Al target for 72\,h with the maximum proton current of 500\,nA available at HZB would yield an $^{22}$Na activity of 92\,kBq in a 50\,$\mu$m thin Al foil.
A cyclotron providing, e.g. 10$\mu$A of 68\,MeV protons would generate 1.2\,MBq of $^{22}$Na in 50\,$\mu$m Al.
In this case, however, a target has to be designed including active cooling to dissipate the high thermal power of 680\,W.


\section{Summary}
In this work, we exploited the $^{27}$Al(p,x)$^{22}$Na reaction for the production of positron emitting Al discs using the 68\,MeV proton beam provided by the cyclotron at the Helmholtz-Zentrum in Berlin. 
Simulations were conducted to investigate the energy loss, the range and the radial scattering of the proton beam in aluminum in order to determine the target dimensions. 
We designed a simple irradiation target comprising a stack of aluminum discs that allowed the straightforward measurement of the depth and lateral distribution of $^{22}$Na.
The activity of other nuclides produced was shown to be negligible after a time of about 15 days.
The comparison of the experimental data with the simulation result based on literature data allowed us to determine new values for the cross section of the $^{27}$Al(p,x)$^{22}$Na reaction.
The  basic concept for $^{22}$Na production in thin Al foils presented here is expected to be promising as potential route for source production in large scale as main advantages are (i) simultaneous production of multiple positron sources at once and (ii) their simple handling since, e.g.\,wet-chemical processes for extraction of $^{22}$Na are avoided. 

\bibliography{main.bib}

\newpage

\begin{appendices}
\section{New Cross Section Values for the $^{27}$Al(p,x)$^{22}$Na Reaction}
\label{appendix:crosssection}

\begin{table}[!htb]
    \begin{minipage}[t]{.5\linewidth}\vspace{0pt}
        \begin{tabular}{l|l}
            E [MeV] & $\sigma$ [mb]\\
\hline
25.83 & 0.00 \\
27.10 & 0.00 \\
28.37 & 3.48 \\
29.59 & 6.55 \\
30.78 & 10.22 \\
31.90 & 14.45 \\
33.01 & 20.32 \\
34.09 & 24.70 \\
35.15 & 28.15 \\
36.18 & 32.63 \\
37.20 & 36.44 \\
38.17 & 38.26 \\
39.11 & 41.64 \\
40.07 & 42.16 \\
40.98 & 42.47 \\
41.90 & 42.43 \\
42.81 & 43.47 \\
43.70 & 43.76 \\
44.56 & 41.60 \\
45.43 & 41.71 \\
46.29 & 41.19 \\
47.12 & 40.07 \\
47.94 & 38.00 \\
48.75 & 36.71 \\
49.56 & 35.25 \\
50.34 & 34.77 \\
        \end{tabular}
    \end{minipage}%
    \begin{minipage}[t]{.5\linewidth}\vspace{0pt}
        \begin{tabular}{l|l}
         E [MeV] & $\sigma$ [mb]\\
\hline
51.10 & 34.91 \\
51.87 & 33.95 \\
52.65 & 31.74 \\
53.40 & 31.17 \\
54.14 & 29.27 \\
54.85 & 28.87 \\
55.62 & 28.61 \\
56.33 & 28.00 \\
57.04 & 27.55 \\
57.74 & 26.71 \\
58.46 & 24.99 \\
59.16 & 25.81 \\
59.85 & 26.14 \\
60.53 & 25.81 \\
61.21 & 24.75 \\
61.85 & 24.74 \\
62.53 & 25.04 \\
63.21 & 25.02 \\
63.86 & 24.35 \\
64.51 & 23.01 \\
65.16 & 23.90 \\
65.80 & 23.00 \\
66.43 & 22.85 \\
67.06 & 23.76 \\
67.67 & 23.14 \\       
        \end{tabular}
    \end{minipage} 
\caption{The values of the cross section are obtained by fitting the experimental data shown in Figure\,\ref{fig:correctedcs}.}
   \label{tab:NEWcrosssection}
\end{table}

\end{appendices}

\end{document}